\documentclass{aa}
\usepackage[english]{babel}
\usepackage{units}
\usepackage{graphicx}
\usepackage{amsmath}
\usepackage{amssymb}
\usepackage{appendix}

\newcommand{\E}{\mathrm{E}}
\newcommand{\e}{\mathrm{e}}

\usepackage{natbib}
\bibpunct{(}{)}{;}{a}{}{,} 

\begin{document}

\title{Timing calibration and spectral cleaning of LOFAR time series data}

\author{
A.~Corstanje\inst{\ref{ru}} \and
S.~Buitink\inst{\ref{br}} \and
J.~E.~Enriquez\inst{\ref{ru}} \and
H.~Falcke\inst{\ref{ru}\and\ref{ni}\and\ref{as}\and\ref{mpifr}} \and
J.~R.~H\"orandel\inst{\ref{ru}\and\ref{ni}} \and
M.~Krause\inst{\ref{desy}} \and
A.~ Nelles\inst{\ref{ru}\and\ref{ca}} \and
J.~P.~Rachen\inst{\ref{ru}} \and
P.~Schellart\inst{\ref{ru}} \and
O.~Scholten\inst{\ref{rug}\and\ref{hebr}} \and
S.~ter Veen\inst{\ref{ru}} \and
S.~Thoudam\inst{\ref{ru}} \and
T.~N.~G.~Trinh\inst{\ref{rug}} 
}

\institute{
Department of Astrophysics/IMAPP, Radboud University Nijmegen, P.O. Box 9010, 6500 GL Nijmegen, The Netherlands\label{ru}\and
Nikhef, Science Park Amsterdam, 1098 XG Amsterdam, The Netherlands\label{ni}\and
Netherlands Institute for Radio Astronomy (ASTRON), Postbus 2, 7990 AA Dwingeloo, The Netherlands\label{as}\and
Max-Planck-Institut f\"{u}r Radioastronomie, Auf dem H\"ugel 69, 53121 Bonn, Germany\label{mpifr}\and
KVI-CART, University Groningen, P.O. Box 72, 9700 AB Groningen, The Netherlands\label{rug}\and
Astrophysical Institute, Vrije Universiteit Brussel, Pleinlaan 2, 1050 Brussels, Belgium\label{br}\and
Department of Physics and Astronomy, University of California Irvine, Irvine, CA 92697-4575, USA\label{ca}\and
Deutsches Elektronen-Synchrotron (DESY), Platanenallee 6, 15738 Zeuthen, Germany\label{desy}\and
Interuniversity Institute for High-Energy, Vrije Universiteit Brussel, Pleinlaan 2, 1050 Brussels, Belgium\label{hebr}
}

\abstract{
We describe a method for spectral cleaning and timing calibration of short voltage time series data from individual radio interferometer receivers.
It makes use of the phase differences in Fast Fourier Transform (FFT) spectra across antenna pairs. For strong, localized terrestrial sources these are stable over time, while being approximately uniform-random for a sum over many sources or for noise. Using only milliseconds-long datasets, the method finds the strongest interfering transmitters, a first-order solution for relative timing calibrations, and faulty data channels. No knowledge of gain response or quiescent noise levels of the receivers is required. 
With relatively small data volumes, this approach is suitable for use in an online system monitoring setup for interferometric arrays. \\
\\
We have applied the method to our cosmic-ray data collection, a collection of measurements of short pulses from extensive air showers, recorded by the LOFAR radio telescope. Per air shower, we have collected $\unit[2]{ms}$ of raw time series data for each receiver.
The spectral cleaning has a calculated optimal sensitivity corresponding to a power signal-to-noise ratio of 0.08 (or -11 dB) in a spectral window of 25 kHz, for 2 ms of data in 48 antennas. This is well sufficient for our application. 
Timing calibration across individual antenna pairs has been performed at $\unit[0.4]{ns}$ precision; for calibration of signal clocks across stations of 48 antennas the precision is $\unit[0.1]{ns}$.
Monitoring differences in timing calibration per antenna pair over the course of the period 2011 to 2015 shows a precision of \unit[0.08]{ns}, which is useful for monitoring and correcting drifts in signal path synchronizations.\\
\\
A cross-check method for timing calibration is presented, using a pulse transmitter carried by a drone flying over the array. Timing precision is similar, \unit[0.3]{ns}, but is limited by transmitter position measurements, while requiring dedicated flights.}


\maketitle

\section{Introduction}
An interferometric radio telescope relies on an accurate timing calibration of the signals of all its constituent receivers, in order to be able to combine signals with a time or phase shift corresponding to the direction of a given source in the sky. Furthermore, spurious narrow-band transmitter signals, which are present even in relatively radio-quiet regions, will show up also in the processed signals. These have to be identified and removed, preferably early in the analysis process.

This paper is organized as follows: in Sect.~\ref{sect:existing} we briefly review some methods that are used for detection and removal of radio-frequency interference (RFI), as well as methods for timing and phase calibration. In Sect.~\ref{sect:thismethod} we introduce our methods; Chapter~\ref{sect:method_detail} describes the methods in detail, and in Chapter~\ref{sect:application}, their application to data taken with the LOFAR radio telescope is discussed.

\subsection{Existing methods for spectral cleaning and timing calibration}\label{sect:existing}
Most of the radio-frequency interference (RFI) present at radio telescope sites consists of either narrow-band signals from radio transmitters, or short pulses in the time domain \citep{Offringa:2013}.
For both cases, there are several methods being used to identify interference, either before or after signal correlation. 
Before correlation in the interferometer, these algorithms typically involve detecting threshold crossings of amplitudes in the time or frequency domain, where the threshold is adapted based on signal properties \citep{Offringa:2010}. 
For instance, a threshold can be calculated by using a median filter, such as used in the Auger Engineering Radio Array (AERA) for radio detection of cosmic rays \citep{Schmidt:2011}. It replaces a sample in time or frequency domain by the median of a number of its neighbours in order to set a threshold. More elaborate techniques, also exploiting correlations of multiple samples crossing the threshold, are found e.g.~in \citet{Offringa:2010}.

Another approach which has been considered for use in the AERA experiment, is described in \citet{Szadkowski:2013}. It uses linear prediction, implemented as a finite impulse response (time domain) filter in FPGAs. This operates online on single receivers and adapts to changes in the interference environment.

After correlation, one can also use adaptive thresholds, then on correlated visibility amplitudes instead of data streams from single receivers. 

Another method is {\it fringe fitting}, which makes use of the fact that most RFI sources are at a fixed position, and therefore produce sinusoidal fringes in visibility data of a fringe-stopping interferometer \citep{Athreya:2009}. These sinusoids are then fitted and removed. 
This latter method has some similarity to the method we present below, which operates on short time series.

Timing and phase calibration in interferometric radio telescopes is typically done based on the principle of self-calibration (\citet{Pearson:1984}; \citet{Taylor:1999}), where one makes use of redundant information in the interferometric data; for instance, there are $N_{\rm ant}(N_{\rm ant}-1)/2$ baselines giving correlated signals, while having only $N_{\rm ant}$ antennas to calibrate. For this method, suitable calibrator sources for which the structure is known, e.g.~point sources, are used as a model for optimizing the calibration. 
The calibration solution can be obtained as a function of frequency, providing a phase calibration for every frequency in the spectrum. The phase calibration at a given frequency equals a timing calibration at the same frequency, taken modulo the wave period.

Moreover, there are methods that also allow to solve for directional dependencies of the calibration. As antennas have a complex gain that has directional dependence, the calibration in general depends on this as well, especially considering differences in gain between antennas.
One of these methods, that is used at LOFAR, is SAGECal \citep{Kazemi:2011}.
A review of similar calibration methods is given e.g.~in~\citet{Wijnholds:2010}.

Alternative approaches involve calibrating on a fixed custom transmitter, such as done by the LOPES cosmic-ray detection experiment \citep{Schroeder:2010}, which yields a timing calibration per antenna for a single, or a few frequencies. 
In our approach, as described below, we use the spectral cleaning method to identify a suitable public transmitter, and also make use of the position of the most useful transmitter in order to obtain a calibration solution.
This is sufficient for a precise (sub-nanosecond) timing analysis of cosmic ray pulses \citep{Corstanje:2015}. It can also serve as a starting point and cross-check for dedicated phase calibrations as used in radio astronomy.

Instead of a fixed transmitter, one could also use satellites or drones flying overhead, with which amplitude calibration is possible as well. This is similar to the amplitude calibration from a fixed transmitter as has been performed at LOFAR \citep{Nelles:2015}. 

Calibration on pulses from the far field, e.g. emitted by airplanes passing overhead, has also been considered \citep{Aab:2016}. However, this relies on randomly occurring pulses that one needs to trigger on in real-time in order to record them. 

\subsection{This analysis}\label{sect:thismethod}
Here, we describe a method of spectral cleaning of time series data that we use to remove narrow-band radio-frequency (RF) transmitter signals from our data. At the same time it allows to obtain a calibration of clock differences across the array.
The method applies only to narrow-band signals, that are present continuously for about 0.2 to $\unit[2]{ms}$, where shorter signals need to be stronger to be detected. Signals with somewhat larger bandwidth are treated as a set of narrow-band signals.
Broadband pulses are not removed. 

Using the phase component of the Fourier transform of each channel, we make use of the fact that strong, localized transmitters produce approximately constant phase differences across the array. Astronomical signals are typically broad-band, and arrive at the antennas as a sum over many sources on the sky, and therefore produce random phase differences over time. This difference allows for an accurate identification (and removal) of disturbing signals. Using the identified constant phases of a public radio transmitter signal, we can also calibrate signal timing offsets in each antenna pair.
If the geometric delay from the signal path lengths of the radio signal to each antenna is known, this leads to a known difference in phase of the (continuous-wave) signal as it is measured at each antenna. Comparing the actually measured phases with the expected phases gives a calibration correction.
It has been suggested as a promising improvement in \citet{Offringa:2010} to add the use of phase information to existing amplitude-based RFI cleaning methods. The method presented here uses only the phase component. 

We apply this method to data taken with the Low Frequency Array (LOFAR) \citep{van-Haarlem:2013} radio telescope. The antennas of LOFAR are distributed over northern Europe, with the densest concentration in the north of the Netherlands. The antennas are organized into stations, each consisting of 96 low-band antennas (LBA, 10 - 90 MHz), and 48 high-band antennas (HBA, 110 - 240 MHz). Within the core region of about $\unit[6]{km^2}$, 24 of these stations have been distributed. 

For the cosmic-ray data collection, we record radio emission from extensive air showers, reaching the ground as a short pulse, on the order of 10 to $\unit[100]{ns}$ long \citep{Schellart:2013}. We use the Transient Buffer Boards installed in the data channel of every LOFAR antenna to record these, as well as other fast radio transients.  
Each recording is 2 to $\unit[5]{ms}$ long and consists of the raw voltage time series from every data channel. The buffer is capable of storing signals up to 5 seconds length. 

These datasets need spectral cleaning in order to measure the pulses accurately. The relative timings of the pulses contain information about the air shower process. For instance, by measuring pulse arrival times, we have evaluated the shape of the radio wavefront as it arrives at the antenna array \citep{Corstanje:2015}.

As our datasets are very short compared to typical astronomical observations (a few milliseconds, instead of hours), and are stored as unprocessed voltage time series per receiver, a dedicated spectral cleaning method is required. Still, our method can be easily adapted for other purposes and instruments, as long as raw time series are available.

We have tested our timing calibration using a pulse transmitter carried by an octocopter drone flying above the array. The precision of the pulse arrival time measurements is similar to the phase measurements.

\section{Method}\label{sect:method_detail}
In this section we explain in detail the method and performance of our RFI identification algorithm, and show how the phases of the thereby identified strong transmitters can be used for timing calibration. 
\subsection{Radio frequency interference identification}\label{sec:rfimethod}
In order to identify frequencies that are contaminated by human-made interference, a typical approach is to search for strong signals above the noise level in an amplitude or power spectrum. However, this requires knowledge of the noise power spectra in the absence of RFI transmitters, or an adaptive or iterative technique to estimate these, as mentioned in Sect.~\ref{sect:existing}.

Therefore, we use the relative phases between pairs of antennas. At the frequency used by a transmitter, the phase difference across an antenna pair is stable over time. After all, the signals are typically transmitted from a fixed location, or effectively fixed, on millisecond timescales. 
In contrast, at frequencies where no terrestrial transmitter is present, we measure emission from the Milky Way as well as electronic noise. The Galactic emission is a sum of many sources, assuming the antennas are omnidirectional or have a substantial field of view. 
Therefore, the detected phases can be treated as random on millisecond integration timescales.

In situations where one localized source in the sky fully dominates the signal, such as e.g.~during strong solar bursts, this assumption is not valid. However, this only happens for a small fraction of the time.

We take phase measurements from a Fast Fourier Transform (FFT) of consecutive data blocks for every antenna. One antenna is taken as reference; for every frequency channel, its phase is subtracted to measure only relative phases. 

It is also possible to consider the phase differences across all antenna pairs (baselines), instead of selecting a single antenna as reference. This is more sensitive (see Sect.~\ref{sect:sensitivity}), but also requires more computation time, and hence can be omitted if the single-reference approach meets the requirements for spectral cleaning.

For every frequency channel we calculate the average and variance of the phase over all data blocks. 
The phase average across antenna indices $j$ and $k$ for frequency $\omega$ is defined as follows (denoting relative phases as $\Phi$ and the data block number as superscript $l$):
\begin{equation}
\Phi_{j,k}^l(\omega) = \phi_j^l(\omega) - \phi_k^l(\omega),
\end{equation}
\begin{equation}\label{eq:phaseaverage}
\bar{\Phi}_{j,k}(\omega) = \arg\left(\sum_{l=0}^{N_{\rm blk}-1} \exp(i\, \Phi_{j,k}^l(\omega))\right),
\end{equation}
and the phase variance as
\begin{equation}\label{eq:phasevariance}
s_{j,k}(\omega) = 1 - \frac{1}{N_{\rm blk}}\left|\sum_{l=0}^{N_{\rm blk}-1} \exp(i\, \Phi^l(\omega))\right|,
\end{equation}
where $\Phi^l(\omega)$ is the relative phase measured in data block $l$ at frequency $\omega$, and $N_{\rm blk}$ is the number of data blocks.
The phase variance $s_{j,k}(\omega)$ is close to unity for completely random phases, and zero for completely aligned phases.

If the phases follow a narrow, peaked distribution around the average, with variance $\sigma^2$, then the phase variance is $s_{j,k} = \frac{\sigma^2}{2} + \mathcal{O}(\sigma^4)$. Hence, this quantity is then indeed proportional to the variance. For wider distributions, the $2\pi$-periodicity of phases becomes important, and the phase variance has a maximum value of unity for a uniform distribution, in the large-$N$ limit.

For random phases, $N_{\rm blk}(1-s_{j,k})$ describes the traveled distance in a two-dimensional random walk, as the right-hand part of Eq.~\ref{eq:phasevariance} represents the length of the sum-vector of $N_{\rm blk}$ unit vectors, each of which having a random direction in the complex plane.

For large $N_{\rm blk}$, this distance follows a Rayleigh distribution with scale parameter $s=\sqrt{N_{\rm blk}/2}$ \citep{Rayleigh:1905}, and has an expectation value of $\alpha\,\sqrt{N_{\rm blk}}$, with $\alpha = \frac{1}{2}\sqrt{\pi}\approx0.89$. It has a standard deviation of $\beta\, \sqrt{N_{\rm blk}}$, with $\beta=\sqrt{1-\frac{\pi}{4}}\approx0.46$.
In practice, the large-$N$ approximation is already accurate for $N_{\rm blk} \gtrsim 10$.

Therefore, we have
\begin{equation}\label{eq:phase_noiselevel}
s_{j,k}(\omega) \approx 1 - \frac{\alpha}{\sqrt{N_{\rm blk}}}.
\end{equation}
It is clear that for a coherent, narrowband signal seen at all antennas, the variance should be rather small. 

In order to determine a threshold for significantly detecting a transmitter, we take the average of the phase variances over all antennas or all baselines.
This leaves one averaged phase-variance spectrum, i.e.~one phase variance per frequency channel.
To take the average is a simple, generic choice; when partial detections are expected, e.g.~in a more sparse array or for very nearby RFI, one could consider the full distribution of the phase variance over the antennas, and test for deviations of random behavior. This is however not pursued here. 

We sort the values of the phase variance over all frequencies, and estimate its standard deviation by taking the 95-percentile value minus the median, which is about $1.65\,\sigma$ for Gaussian noise. 
This has the advantage of considering only the upper half of the sample which is assumed to follow the random-walk characteristics. It selects out all transmitter signals, that only lower the variance below the median. This naturally assumes that less than half of the frequency channels contain an interfering transmitter signal, which is reasonable for astronomical observations in general. Should this not be the case for the particular site, one could take a higher percentile value instead of the median. Alternatively, one could choose to follow directly the random-walk statistics for mean and standard deviation, and compare with the data afterwards.

The threshold is set to the median value minus a multiple of the standard deviation, which is tunable to trade e.g.~a lower false-positive probability for a lower sensitivity.

For the run-time complexity, it is noted that the algorithm requires $N_{\rm blk}\; N_{\rm ant}$ FFTs of a fixed length, set by the desired spectral resolution. Moreover, when treating all baselines, it requires $\mathcal{O}(N_{\rm ant}^2)$ phase spectrum comparisons. 
When instead using a single antenna as reference (or a fixed number of them), only $\mathcal{O}(N_{\rm ant})$ comparisons are done, and the FFTs always dominate.

\subsection{Sensitivity of RFI detection}\label{sect:sensitivity}
Noting the correspondence of the detection of an RFI transmitter to the detection of bias, i.e.~a preference towards a certain direction, in a set of identically distributed random walks, the sensitivity of RFI detection can be analyzed.
The full analysis is deferred to the Appendix. 

We start by assuming a signal is present in the noisy time series of each receiver, with power signal-to-noise ratio $S^2$ defined by the absolute-square of the FFT in one frequency channel, for the signal and the noise respectively. 

With this definition, the sensitivity becomes (asymptotic approximation, see Appendix):
\begin{equation}\label{eq:sensitivity}
S^2 > 3.8 \;\sqrt{k/6}\; N_{\rm blk}^{-1/2}\; N_{\rm ant}^{-1/2},
\end{equation}
aimed at a $k=6$-sigma detection threshold as used in the LOFAR cosmic-ray analysis \citep{Schellart:2013}.
Typical numbers for this analysis are $N_{\rm blk} = 50$ and $N_{\rm ant} = 48$, leading to a threshold of $S^2 = 0.08$, or $\unit[-11]{dB}$, which is easily sufficient for the purpose of analyzing pulses from air showers. With a specific bandwidth fraction in mind, e.g.~for wideband radio signals, this sensitivity can be used to give an upper bound to residual RFI levels.

To put this result into perspective, we consider a straightforward, simplified method, that detects excess power in a power spectrum, averaged over all antennas and data blocks. 
The noise power in a given FFT channel has an exponential distribution \citep{Papoulis:2002}, with mean and standard deviation equal to the mean noise power per channel.
The signal power is uncorrelated to the noise, and hence the total power is the sum of signal and noise power.
Summing spectra of $N_{\rm ant}$ antennas each having $N_{\rm blk}$ blocks then yields a threshold
\begin{equation}\label{eq:asympt_powerexcess}
S^2 = 6\;(k/6)\; N_{\rm blk}^{-1/2}\; N_{\rm ant}^{-1/2},
\end{equation}
plus the uncertainty in determining the average quiescent noise spectrum, which is not flat in general. The asymptotic behavior is therefore the same as in Eq.~\ref{eq:sensitivity}.
It is assumed that estimating a single, flat noise level as for the phase variance (Eq.~\ref{eq:phase_noiselevel}) has a lower uncertainty than estimating a spectrum curve. 

Note that, since both methods use averaging over many blocks, or many phase variance values respectively, by the Central Limit Theorem these averages can be regarded as estimating the mean of a Gaussian distribution. Hence, in both cases the $k-$sigma thresholds refer to exceeding probabilities, and corresponding false alarm rates, of a Gaussian distribution.

Our method based on phases has a somewhat favorable detection threshold, the difference with respect to Eq.~\ref{eq:asympt_powerexcess} being at least $\unit[2.0]{dB}$.
Moreover, it does not require an estimate of the noise spectrum in the absence of transmitters. This has made it easier for us to implement in practice, where background levels are variable.
However, this does not imply that this comparison holds when looking at more elaborate, amplitude-based cleaning methods.

Note that when, instead of power excess, one would use the amplitude excess in an absolute spectrum rather than absolute-squared, the asymptotics of $S$ are the same, i.e.~given by Eq.~\ref{eq:asympt_powerexcess}. The difference is at most $\unit[7]{\%}$ in the constant factor, in favor of the absolute spectrum, in the weak-field and large-$N$ limit.

\subsection{Timing calibration}
Observations using an interferometric telescope require precise timing and phase calibration of each receiver, in order to have precise pointings, and to perform imaging with optimal signal quality. The timing precision should be about an order of magnitude below the sampling period.

For timing calibration we use one or multiple narrow-band transmitters as a beacon, producing fixed relative phases between antennas at the transmitting frequency. This is similar to the procedure followed in \citet{Schroeder:2010}; we extend this by a more precise phase measurement, and by using the geometric delays from the transmitter location to find the calibration delays per antenna pair.

We measure relative phases per antenna pair in the same way as in Sect.~\ref{sec:rfimethod}, taking Fourier transforms of consecutive data blocks for all antennas, and averaging phases using Eq.~\ref{eq:phaseaverage}. This also allows to identify frequencies suitable for timing calibration from the values of the phase variance, Eq.~\ref{eq:phasevariance}, where lower values are better. 

The geometric delays from the transmitter to each antenna are needed for determining the calibration delays between antennas.
Therefore, it is required to use a transmitter at a known location, and with frequency above about \unit[30]{MHz}. At lower frequencies, i.e.~the HF-band, one may have signals reflecting off the ionosphere, and propagation characteristics may vary from time to time; see e.g.~\citet{Gilliland:1938}.

The signals propagate with the light speed in air, which is $c/n$. The index of refraction $n$ is on average $1.00031$, noting that variations of $\pm\, 4\,10^{-5}$ are possible with temperature and humidity \citep{Grabner:2011}. Omitting the refractive index would therefore introduce a timing mismatch of $\unit[1.0]{ns}$ between two antennas separated by $\unit[1]{km}$, along the line-of-sight to the transmitter. This is therefore significant at intermediate and longer baselines.

The phase difference across a given antenna pair, after accounting for geometric delays from the transmitter, corresponds to a time difference (calibration mismatch) 
\begin{equation}\label{eq:time_phase}
\Delta t = \frac{\Delta \phi}{2\pi f}\;\;({\rm mod}\;\frac{1}{f}).
\end{equation}
Thus, the calibration solution obtained by using one transmitter is only determined up to a multiple of the signal period $1/f$. This can be improved by combining results from multiple transmitters. However, in order to obtain the correct solution, it is then required that the different transmitters have large differences in period compared to the phase / timing noise. Moreover, in general the correct calibration phase depends on frequency, i.e.~the optimal phase calibration may have deviations from the group delay, as a function of frequency. 
When using a custom beacon for calibration measurements as described here, one would therefore choose frequencies far apart.

Once antenna timings have been calibrated, the relative phases can be monitored over time without reference to the transmitter location and geometric delays.

\section{Application to LOFAR data}\label{sect:application}
In this section, we describe how the RFI identification and timing calibration methods are used for the analysis of air shower datasets with LOFAR.
\subsection{RFI identification}
The LOFAR radio telescope, located in the north of the Netherlands, is in a relatively radio-quiet region. Nevertheless, in all observations there are several signals present, coming from narrow-band transmitters. Therefore, spectral cleaning methods are required to remove them from astronomical observation data. 

\begin{figure}
\begin{center}
\includegraphics[width=0.50\textwidth]{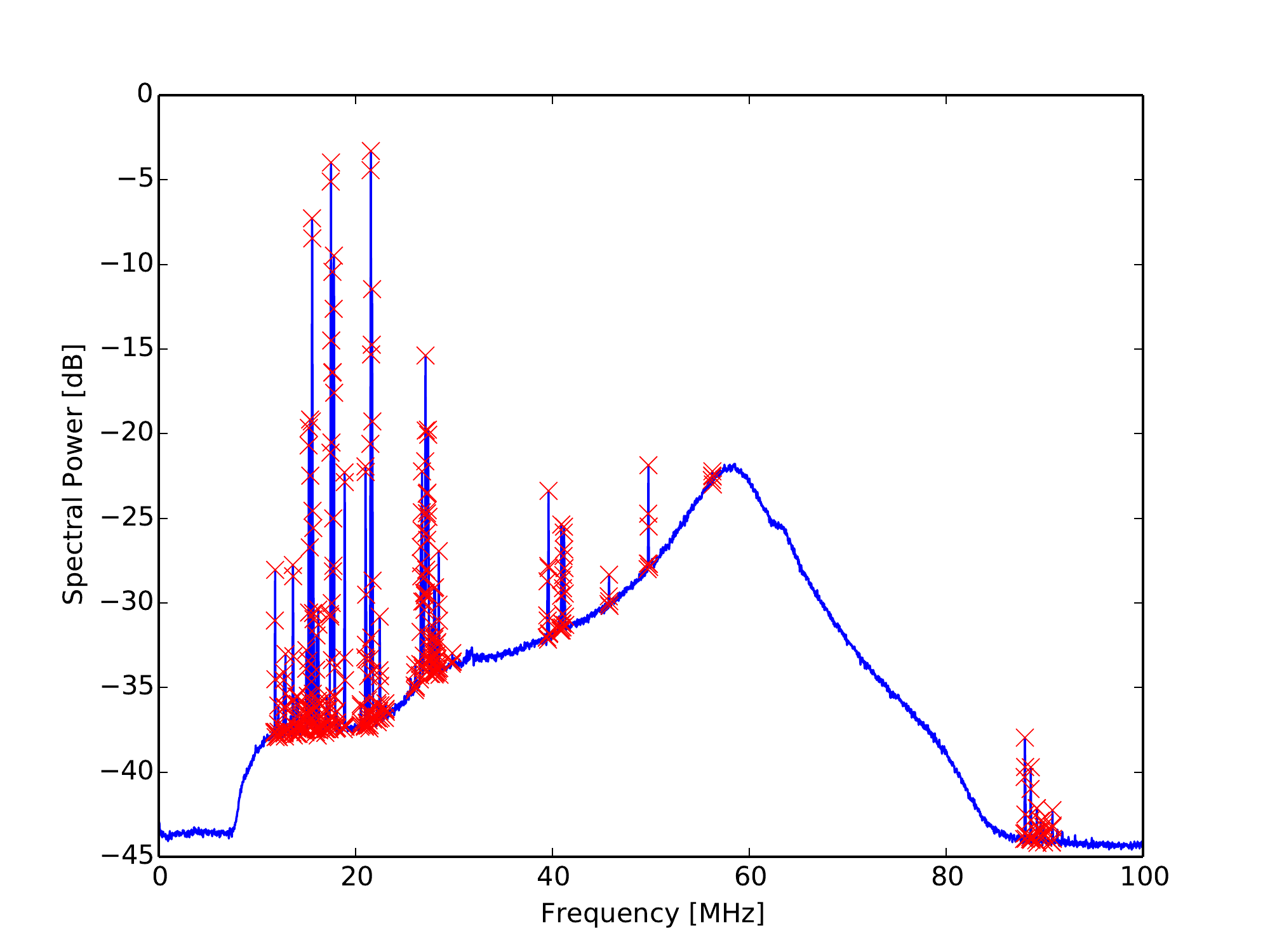}
\caption{Example power spectrum from $\unit[2]{ms}$ of LOFAR data, averaged over 48 LBA antenna dipoles, with crosses indicating channels with detected transmitters.}
\label{fig:flagged_channels}
\end{center}
\end{figure}

\begin{figure}
\begin{center}
\includegraphics[width=0.50\textwidth]{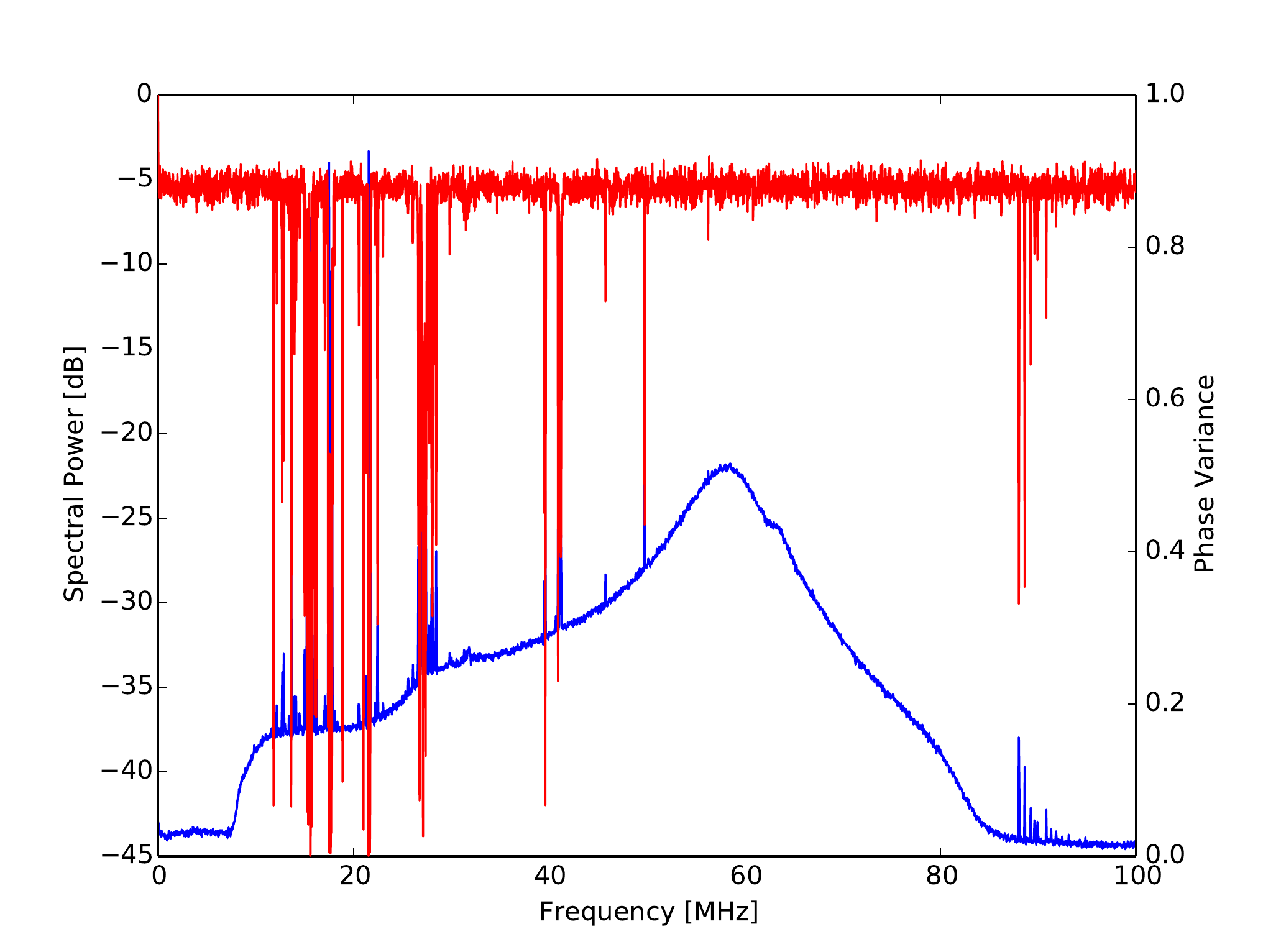}
\caption{Example power spectrum from $\unit[2]{ms}$ of LOFAR time series data (lower curve). The phase variance is shown in the upper (red) data points. It consistently becomes lower whenever a narrowband transmitter is seen in the power spectrum.}
\label{fig:phasevariance}
\end{center}
\end{figure}

We have used the core stations of LOFAR. For all but a few very bright air showers, our data contain antenna baselines up to about $\unit[1]{km}$, and the majority of antennas with signal is in the central ring of \unit[320]{m} diameter.
An example power spectrum is shown in Fig.~\ref{fig:flagged_channels}. 
For demonstration purposes, the dataset of this example has particularly bad RFI, as there are several flagged frequencies in the 30 to $\unit[80]{MHz}$-band. However, this case is still not too extreme, and spectral cleaning is indeed necessary in similar instances.
The power spectrum is averaged over the 48 antennas in one LOFAR station, and averaged over $\unit[2]{ms}$, being the length of a typical cosmic-ray dataset. We treat each of the two instrumental polarizations separately, as RFI signals may be detectable in only one of the two polarizations.
It is a spectrum of the LBA antennas, ranging from 10 to 90 MHz. In what follows we focus only on the low-band spectra as these are best used for air shower measurements; the methods work identically for the high-band antenna data.
For the detection of the transmitter frequencies, we use FFTs with a block size of 8000 samples, which amounts to a spectral resolution of $\unit[25]{kHz}$. There are then 50 blocks in a time series of $\unit[2]{ms}$, which are used to calculate the phase variance over the entire $\unit[2]{ms}$ of data as from \ref{eq:phasevariance}. 
The result is shown in Fig.~\ref{fig:phasevariance}. 
The phase variance, taken as the median value of the 48 antennas, is shown as the upper (red) signal. It has random `noise' due to the finite number of data blocks; at frequencies where a narrow-band transmitter is present in the power spectrum (lower curve), the variance is significantly lower. The random noise has a median value of $0.879$, consistent with the expected value from Eq.~\ref{eq:phase_noiselevel} of $0.875$ for 50 data blocks. This is a basic test of our randomness assumption for the phases. 

The phase variance threshold is then set to a value of (nearly) 6 sigma. The standard deviation is estimated by the 95-percentile value minus the median, which is about $1.65\,\sigma$ for Gaussian noise. 
Every frequency channel with lower phase variance is flagged.

When frequency resolution (set by the chosen FFT block size) is high enough to resolve the transmitters' frequency responses, it can be necessary to also flag a number of adjacent frequency channels, as the edges of resolved transmitter spectra may not meet the threshold criterium for flagging.  This is especially important when a large block size is taken, e.g.~to comply with FFT resolution used in later analysis.
The number of adjacent channels to flag, is currently set as a manually tunable parameter, scaling with frequency resolution. 
\begin{figure}
\begin{center}
\includegraphics[width=0.50\textwidth]{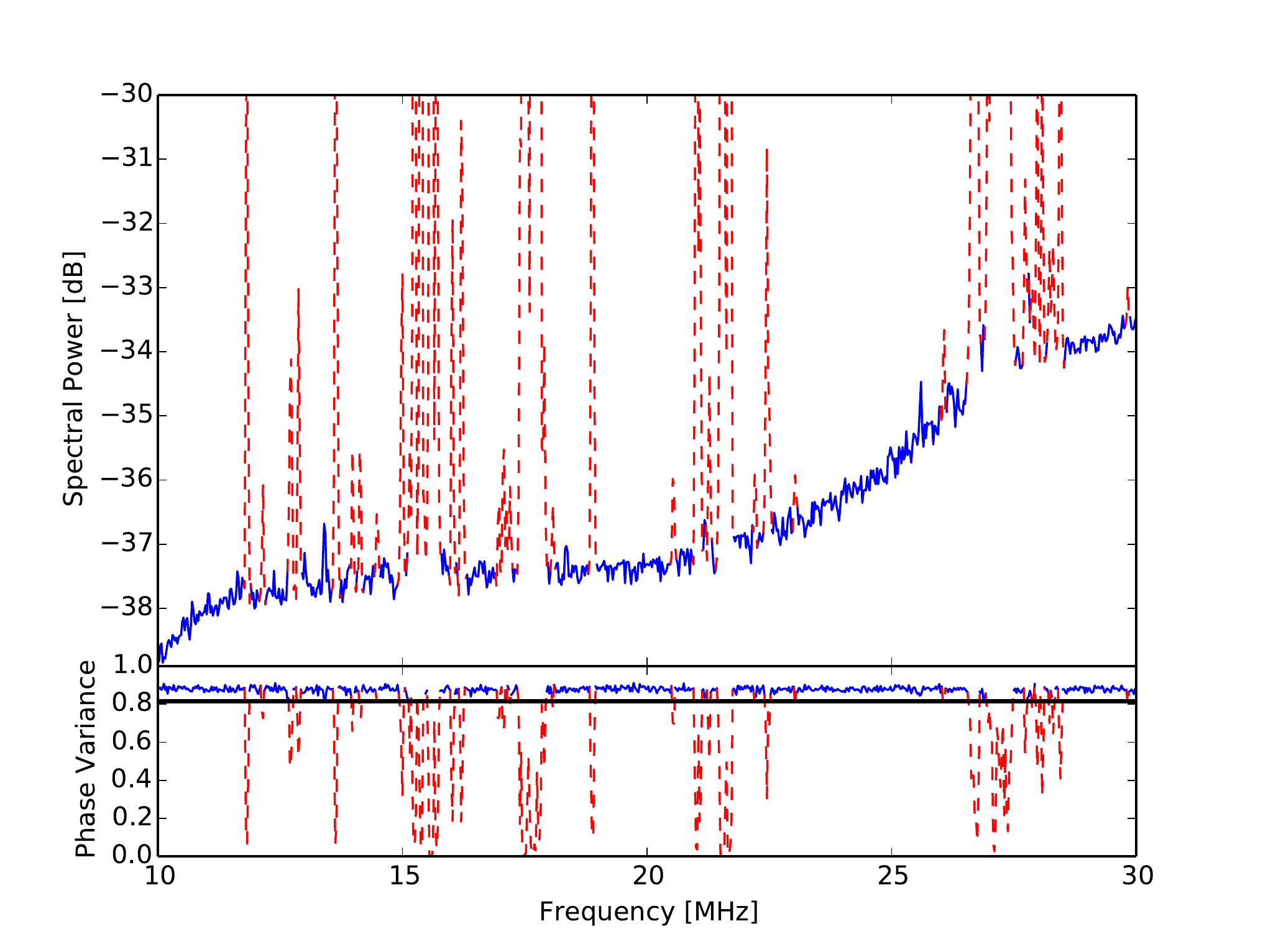}
\caption{Close-up of the power spectrum in a frequency range with several RFI sources. 
Flagged frequencies are shown as red dashed lines; the lower panel shows the phase variance, with the black horizontal line denoting the threshold for flagging.
Although the RFI-quiet noise level would follow a smooth curve, fitting the curve and RFI flagging using the excess power are interdependent. }
\label{fig:spectrum_closeup}
\end{center}
\end{figure}

In Fig.~\ref{fig:spectrum_closeup}, a close-up of the power spectrum and the phase variance are shown. 

\subsection{Timing calibration: results for the LOFAR core}\label{sect:timing_calibration_results}
For calibration of short time series, i.e.~2 to $\unit[5]{ms}$ length, we use one or multiple narrow-band transmitters as a beacon, producing fixed relative phases between antennas at the transmitting frequency. 
The signals at the high end of the spectrum ($> \unit[87]{MHz}$) are from public radio transmitters which are always present. They are well detectable despite being outside the passband of the filters, which ends at $\unit[80]{MHz}$. 
Moreover, the phase variance we measure in the spectral cleaning algorithm, and the corresponding timing precision, is found to be best for these frequencies. 
Therefore, we work with the high-frequency transmitters, especially the strongest one at $\unit[88.0]{MHz}$. 

The radio signals at frequencies 88.0, 88.6, 90.8 and 94.8 MHz are transmitted from a 300 meter high radio tower located in Smilde\footnote{GPS coordinates: $\unit[6.403565]{^\circ}$ East, $\unit[52.902671]{^\circ}$ North.}, at $\unit[31.8]{km}$ from the LOFAR core. 

For $\unit[88.0]{MHz}$, the signal period is $\unit[11.3]{ns}$ which is still large compared to the desired (and attainable) sub-nanosecond calibration precision.

The timing calibration signal follows from the relative phases after accounting for the geometric delays between transmitter and antennas, according to Eq.~\ref{eq:time_phase}.
The relative phases are once again obtained from the FFT of 50 consecutive data blocks, taking average phases as from Eq.~\ref{eq:phaseaverage}. 
As was done for the RFI detection method, we treat the two polarizations of the LOFAR LBA antennas separately. 
We thereby make use of the identical design and orientation of the LOFAR antennas. If antenna orientations or the design of their polarizations are different, this could lead to larger timing errors in this procedure, when using transmitters with polarized signals.
Monitoring of a (cross-)calibration over time would still be accurate, see Sect.~\ref{sect:monitoring} below.

The geometric delays are calculated using the International Terrestrial Reference Frame (ITRF) coordinates \citep{Altamimi:2002} of each antenna, and the GPS (WGS-84) \citep{DMA:1987} ellipsoid coordinates of the Smilde tower converted to ITRF. This is a cartesian coordinate system, allowing for an easy calculation of straight-line distance between two points.

For the effective height of the emission we consider half the height of the tower; the uncertainty in relative timings per $\unit[100]{m}$ of height, is less than \unit[0.05]{ns} across LOFAR core stations, and below \unit[0.005]{ns} within one station, and therefore negligible for our purposes.  

As a starting point we take an existing LOFAR timing calibration per antenna, which is performed using astronomical phase-calibration a few times a year \citep{van-Haarlem:2013}. 
We compare measured relative phases with those from the straight-line propagation, in the LOFAR core area, consisting of a circular-shaped area of $\unit[320]{m}$ diameter, plus some additional stations up to about $\unit[1]{km}$ away. There are many more stations, but our air shower measurements are limited to this area.

\begin{figure}
\begin{center} 
\includegraphics[width=0.50\textwidth]{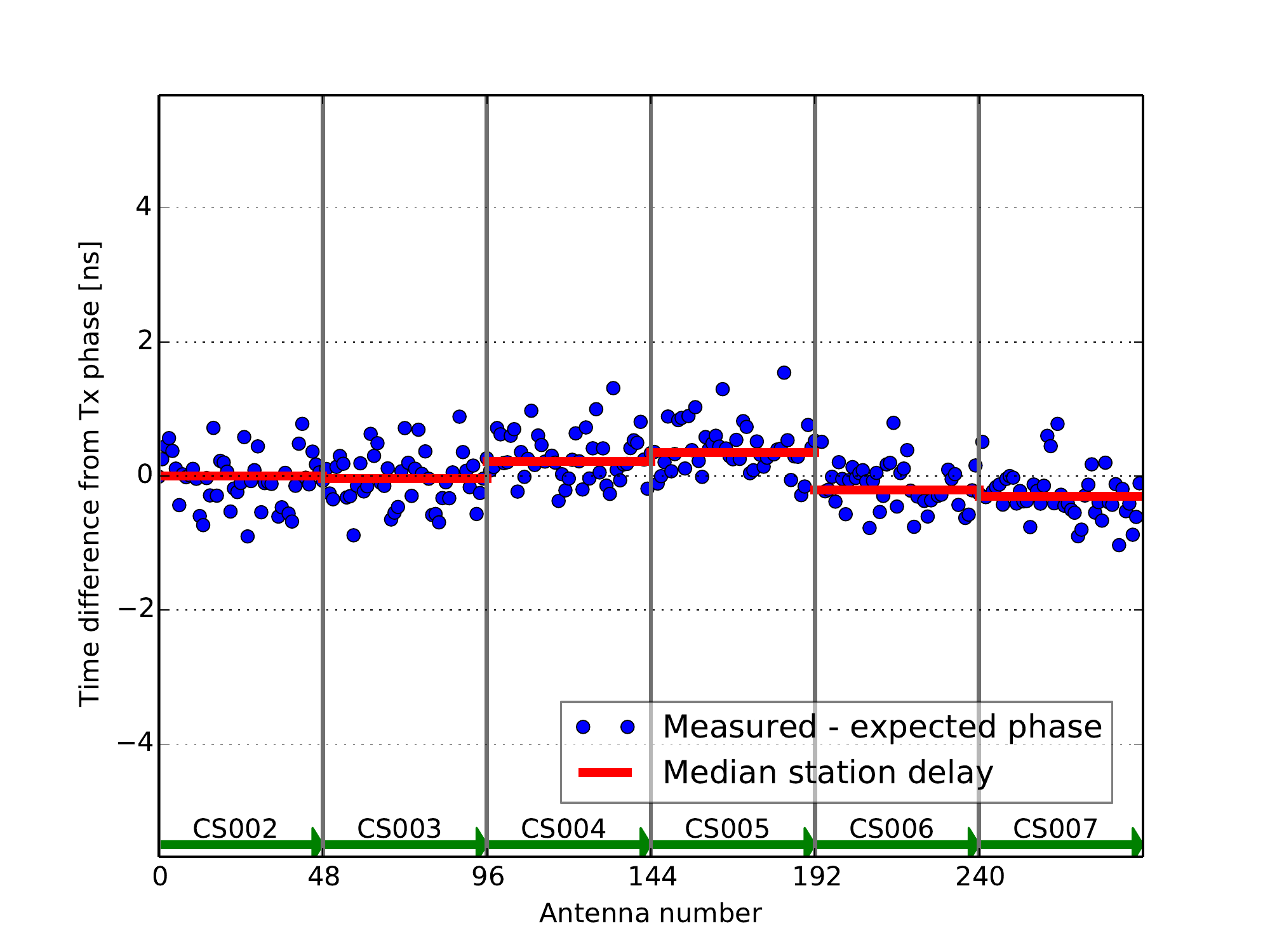}
\caption{Difference between measured and expected phases per antenna, converted to time in nanoseconds. Red (solid) bars represent the median time delay per LOFAR station (48 antennas); the stations are separated by the vertical grid lines. The range of the y-axis corresponds to the signal period at $\unit[88]{MHz}$.}
\label{fig:phases_timing_example}
\end{center}
\end{figure}

The phases correspond to a timing correction per antenna as shown in Fig.~\ref{fig:phases_timing_example}.
The values depicted in this plot consist of both calibration errors and possible systematic effects from our measurement. The latter may include differences in filter characteristics at $\unit[88.0]{MHz}$, i.e.~the delays obtained from phases at this frequency may deviate from the full group delay. Wave propagation effects may vary slightly over antennas, e.g.~due to the presence of other LOFAR antenna(s) along the line of sight to the transmitter. 

We can assume that any calibration mismatch with respect to the earlier LOFAR calibration is independent from these systematic effects. 
Dedicated calibration observations use astronomical sources instead of a terrestrial transmitter, and span the entire frequency band.

Important to note, therefore, is that the timing correction signal we find here provides an upper limit on both calibration errors and systematic effects.

The standard deviation of the timing correction signal is $\unit[0.44]{ns}$. Per station, the standard deviation varies from $0.36$ to $\unit[0.40]{ns}$. 

Our measurements and data taking have started in June 2011, which was within the commissioning period of LOFAR; the `cycle 0' observations have started in December, 2012. This means that some technical timing issues that have been resolved later, were still there. Using this method, these have been detected and corrected, from the same datasets that contain our cosmic-ray measurements. Hence, also our older data can be fully used.

As LOFAR is divided into separate stations, timing calibration across stations is also required.
Especially before October 2012, only the six innermost stations had a common clock, but all other core stations had their own clock synchronized by GPS. This caused clock drifting across stations on the order of $\unit[10]{ns}$, which is long compared to interferometric accuracy requirements. 

Therefore, we calculate the inter-station clock offsets by taking the median of the time delays per antenna in each station. 
Using the median instead of the mean is more robust against calibration errors or malfunctioning of a small fraction of antennas. 
On the other hand, the median has a higher uncertainty for estimating the mean than taking the average. Still, taking the median is useful when batch-processing thousands of datasets.

When inter-station clock offsets vary by more than the signal period of $\unit[11.3]{ns}$, they are still known accurately up to a multiple of this period. For the cosmic-ray pulse timing measurements as performed in \citet{Corstanje:2015}, the actual solution can be identified by using fits of the incoming direction of the radio pulse of the air shower. These fits are done on single-station level and hence are not influenced by the inter-station offsets.

The standard error of the median over one station amounts to $\unit[0.08]{ns}$, and is a factor $\sqrt{\pi / 2} \approx 1.25$ higher than the standard error of the mean.
Therefore, the inter-station clock offsets can be determined to about $\unit[0.1]{ns}$ precision, assuming systematic effects average out over the antennas of each station.

\subsubsection{Multiple transmitters for calibration}
The calibration solution obtained from using one transmitter is only given up to a multiple of the signal period. This can be improved by combining results from multiple transmitters. However, to obtain the correct solution, it is required that the different transmitters have large differences in period compared to the phase / timing noise. For the LOFAR environment, the difference in period between 88.0 and $\unit[90.8]{MHz}$ is only $\unit[0.35]{ns}$ which is not always above the timing noise. The transmitter at $\unit[94.8]{MHz}$ is not as reliable as its signal is rather weak. 
Moreover, in general the correct calibration phase depends on frequency, i.e.~the optimal phase calibration may have deviations from the group delay, as a function of frequency. This leads to an additional source of uncertainty when combining multiple frequencies.

When instead using a custom beacon for calibration measurements, in the way we described here for the public radio signals, one would choose frequencies further apart. Also in this case, differences in phase delay versus group delay may show up.
One could as well use a beacon sending short pulses or bursts, as we show in the next section. These pulses do not have issues with periodicity.

\subsubsection{Pulse arrival times from an octocopter drone}\label{sect:octocopter}
As a cross-check, we have performed a pulse arrival time measurement in the LOFAR inner core region, using a pulse transmitter mounted below an octocopter drone. The octocopter flies with a pre-programmed flight path, using GPS coordinates. 
We have set it to fly above the central antenna of the 6 innermost stations of LOFAR. A pulse of approximately $\unit[250]{V}$ is then transmitted every $\unit[8]{\mu s}$ from a height of about $\unit[50]{m}$.
The incoming signal is recorded using the Transient Buffer Boards. The individual pulses are timed by interpolating the time series using up-sampling, and taking the time of the first positive maximum after the signal exceeds a given threshold, set as a fraction of its amplitude. This method was found suitable for timing relatively long pulses with a broad maximum. The rise time of the pulses was on the order of $\unit[50]{ns}$, corresponding to about 3 periods at the resonance frequency of the LBA antennas, near $\unit[58]{MHz}$. The pulses showed a strong signal-to-noise ratio in all antennas we used, hence it was possible to identify the correct maximum for timing. 

Geometric delays follow from a straight-line path from the pulse transmitter to each antenna; the calibration signal for each antenna pair is the remaining time delay after accounting for the geometric delays.

The actual position of the octocopter can vary due to wind and flight control uncertainties. In order to determine the transmitter location more precisely at the time of the measurement, an optimization procedure has been performed. The calibration signals have been minimized with respect to a given calibration of  LOFAR, which for the majority of antennas has an uncertainty of at most $\sigma = \unit[0.4]{ns}$ as shown in Sect.~\ref{sect:timing_calibration_results}.
The position shifts by the optimization procedure were found to be about 1 to \unit[1.5]{m}, which is significant for timing purposes, when calibrating from scratch. The fit uncertainty then depends nontrivially on the calibration delays themselves.

\begin{figure}
\begin{center} 
\includegraphics[width=0.50\textwidth]{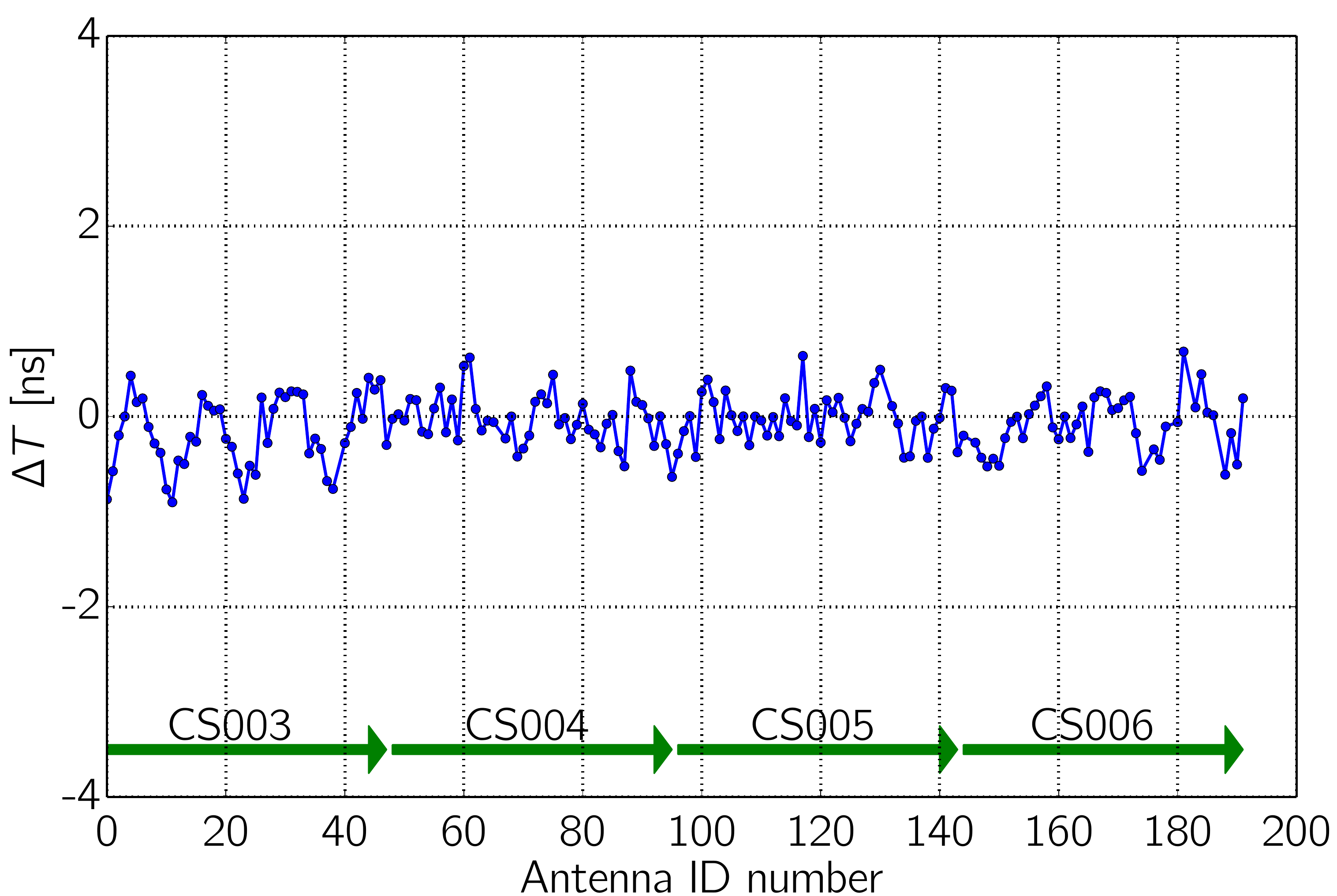}
\caption{Arrival times of pulses from octocopter drone. The points show the difference between measured and expected pulse arrival time, per antenna, for 4 of the innermost LOFAR stations indicated by the labeled arrows at the bottom.}
\label{fig:octocopter_timing_example}
\end{center}
\end{figure}

Comparing pulse arrival times at each antenna with the expected geometric delay of the signal path from transmitter to receiver, we obtain the calibration signal as in Fig.~\ref{fig:octocopter_timing_example}. The calibration signal is an average over 10 pulses.
The standard deviation of the timing calibration signal amounts to \unit[0.26]{ns}.
This is comparable to the result of \unit[0.44]{ns} obtained using continuous-wave radio transmitters. 
Nevertheless, there is still some structure visible in the arrival times for one of the stations, labeled CS003. This may point to a non-optimal fit for the transmitter position.

\subsection{System monitoring}\label{sect:monitoring}
We have monitored the relative delays between antennas over the course of nearly 4 years, comparing the results of the given procedure for all datasets in our collection.
With at least one calibration at a given date, for which we also know the relative phases, the time variations can be monitored without reference to the transmitter location, wave propagation etc. Only the measured relative phases need to be compared. 

\begin{figure}
\begin{center} 
\includegraphics[width=0.50\textwidth]{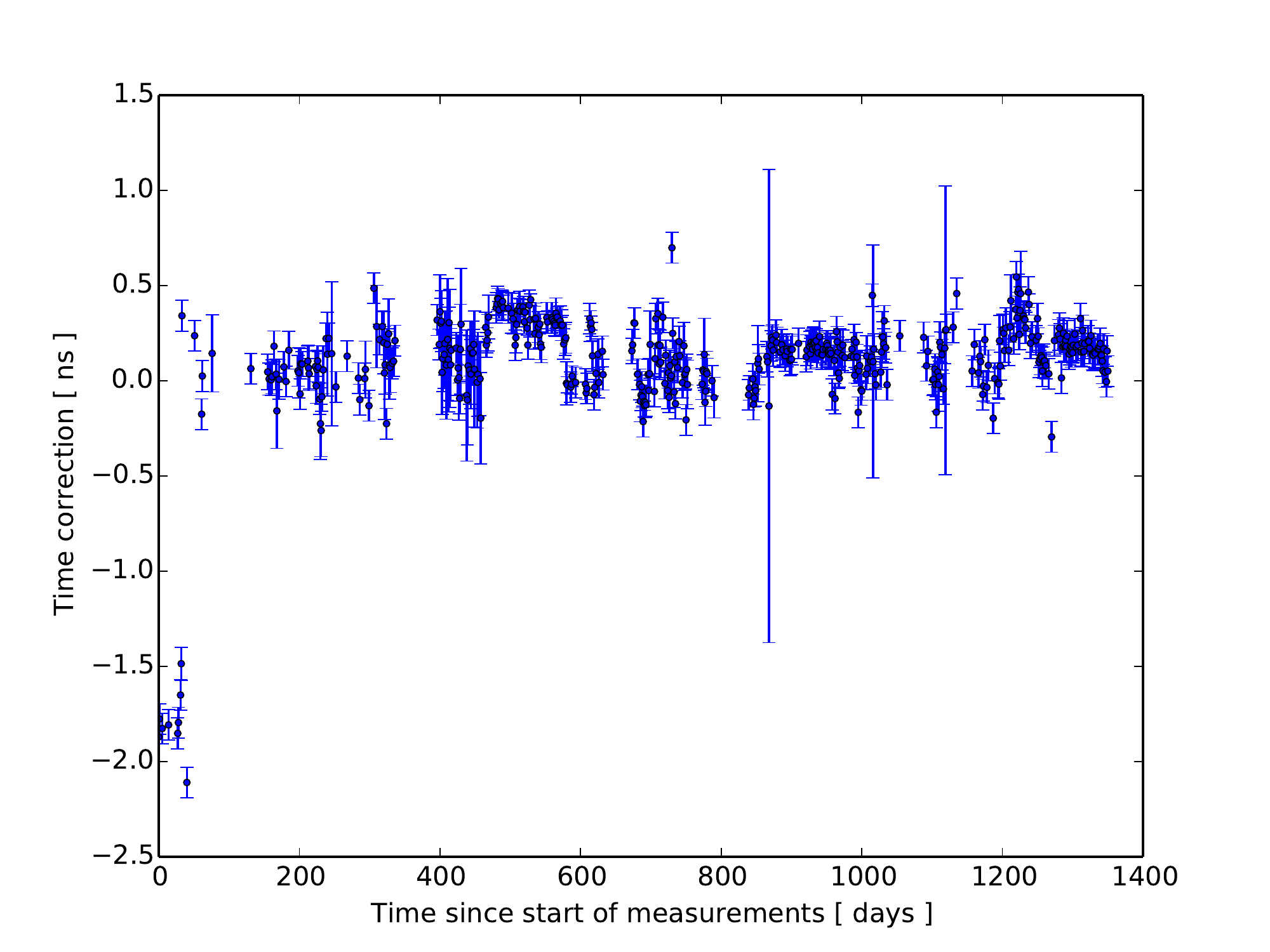}
\caption{Time variation of the relative delay between two antennas within one LOFAR station over the course of our nearly four-year data collection. Residual delay values are binned per day, showing average and standard deviation within one bin.}
\label{fig:timevariation_overall}
\end{center}
\end{figure}

\begin{figure}
\begin{center} 
\includegraphics[width=0.50\textwidth]{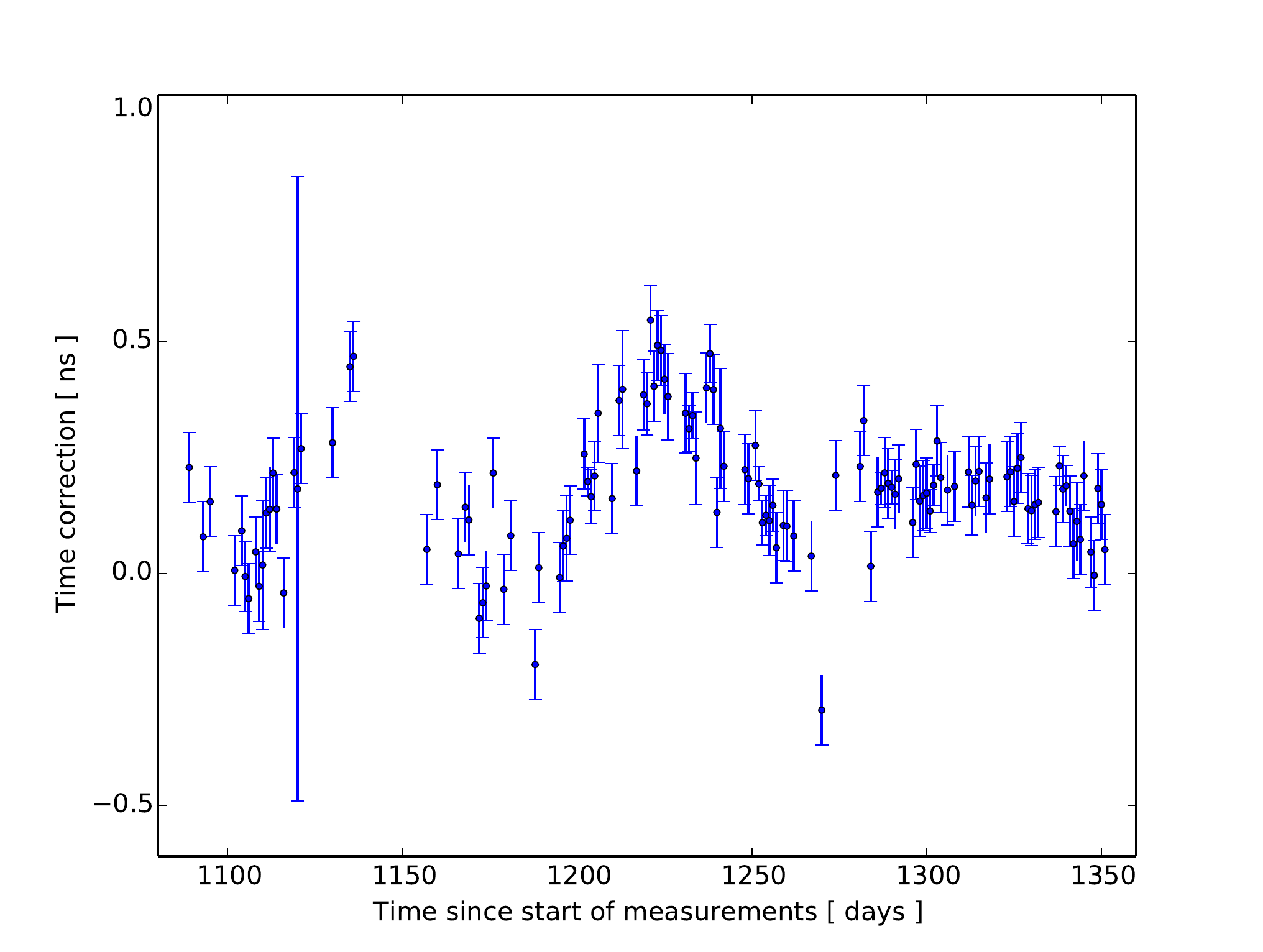}
\caption{A close-up of the time variation of the relative delay for the same antenna pair, showing the precision of the delay monitoring as well as some clock drifting.}
\label{fig:timevariation_closeup}
\end{center}
\end{figure}

A typical time variation plot is given in Fig.~\ref{fig:timevariation_overall}. Timing corrections have been binned, using one bin per day. The given uncertainties are the standard deviations over one day. The median value of this uncertainty is $\unit[0.08]{ns}$, taken only from those days where at least 5 measurements were taken. This median uncertainty is also assigned to data points from days with less than five measurements.
The relative timing between these two antennas is mostly stable over time at the \unit[0.5]{ns} level, except for the first month of measurements which was within the commissioning time of LOFAR.
After this, only on three days there was no stable solution for the timing, showing as large uncertainties in Figs.~\ref{fig:timevariation_overall} and \ref{fig:timevariation_closeup}.

Fig.~\ref{fig:timevariation_closeup} shows a close-up of the same plot. It shows a slow clock drifting, and demonstrates that indeed signal path synchronization at the level of $\unit[0.1]{ns}$ can be followed and corrected.

\section{Conclusion and Outlook}
We have developed a spectral cleaning method and a timing calibration method for interferometric radio antenna arrays. These have been designed to operate on milliseconds-long time series datasets for individual receivers. The methods have been used for our analysis of cosmic-ray datasets, to calibrate and clean voltage time series data. 
Using phases from an FFT for spectral cleaning has shown to be simpler to use than a straightforward threshold in an averaged power spectrum, as no a priori knowledge of the antenna gain curve or noise spectrum is required. 
Moreover, when compared to this average spectrum threshold, the method has a slightly favorable detection power threshold which is at least $\unit[2.0]{dB}$ lower.
In our application, the threshold of the method is at a power signal-to-noise ratio of $\unit[-11]{dB}$ in a $\unit[25]{kHz}$ spectral window.

Timing calibration using the phases of public radio transmitter signals has been performed to a precision of \unit[0.4]{ns} for each antenna, at a sampling period of \unit[5]{ns} (or \unit[200]{MHz} sampling rate). Monitoring a given calibration over time has a precision of \unit[0.08]{ns} for each antenna pair in a LOFAR station.
Obtaining a timing calibration from a pulse transmitter aboard a drone flying over the array is possible to a similar precision of \unit[0.3]{ns}, mainly limited by the accuracy of the position measurement of the transmitter. 

As the methods described here only require datasets with lengths of 2 to $\unit[5]{ms}$, they would be well suited for system monitoring and (pre-)calibration purposes of interferometric radio arrays in general. Apart from detecting interference and timing calibration, one can identify malfunctioning receiver data channels. Examples include zero or unusual signal power, unstable timing calibrations, polarization errors, and outlying receiver gain curves. 
Detecting these issues in an early stage prevents the propagation of faulty signals into the correlation and imaging process, where they are more difficult to remove.

It is expected that future low-frequency radio telescopes such as the SKA-Low (low-frequency part of the Square Kilometre Array) will also be built out of many individual antenna elements laid out in a relatively dense pattern on the ground. In \citet{Dewdney:2015} it is shown that the majority of antennas is planned to be located at a distance of up to $\unit[10]{km}$ from a central core. These would be in the line of sight of a single transmitting beacon, either custom or RFI. Ideally one would use a custom beacon that is turned on only a few parts per million of the time, for calibration. 

Timing and phase calibration of all signal paths is a similar challenge as in LOFAR, only on a much larger scale. 
Even with the use of one common clock signal, the entire signal path to the analog-digital conversion unit can exhibit nontrivial variations over time, e.g.~along the analog signal transport to the central processing facility.
This is already seen in Fig.~\ref{fig:timevariation_closeup}, where the given antenna pair was located inside one LOFAR station, sharing the same clock signal.
The techniques presented here, when merged with more elaborate existing methods, could prove useful for this.

\section{Acknowledgements}
The LOFAR Key Science Project Cosmic Rays greatly acknowledges the scientific and technical support from ASTRON.

The authors thank K.~Weidenhaupt, R.~Krause and M.~Erdmann for providing and operating the octocopter drone.
We also thank the anonymous referee for useful comments.

The project acknowledges funding from an Advanced Grant of the European Research Council (FP/2007-2013) / ERC Grant Agreement n. 227610. The project has also received funding from the European Research Council (ERC) under the European Union's Horizon 2020 research and innovation programme (grant agreement No 640130).   We furthermore acknowledge financial support from FOM, (FOM-project 12PR304) and NWO (VENI grant 639-041-130). AN is supported by the DFG (research fellowship NE 2031/1-1).  

LOFAR, the Low Frequency Array designed and constructed by ASTRON, has facilities in several countries, that are owned by various parties (each with their own funding sources), and that are collectively operated by the International LOFAR Telescope (ILT) foundation under a joint scientific policy.

\begin{appendix}
\section{}
Here we describe the details of the sensitivity analysis for the spectral cleaning method described in Sect.~\ref{sec:rfimethod}.

The problem of finding the threshold for detecting a transmitter can be described as to determine if a given random walk (or ensemble of random walks) is biased or not. The sum of a sequence of phase vectors $\e^{i\,\phi_j}$ forms a random walk in the complex plane, with unit step size.
The random walk is biased if it has a preference towards a certain direction; on average this gives a longer distance for the random walk.

Assume a transmitter signal measured in one frequency channel of the FFT of a noisy time series, with amplitude $a$ at each receiver.
Let the mean noise power in this channel be $\sigma^2$, so the power signal-to-noise ratio is defined as $S^2 \equiv a^2 / \sigma^2$. For this calculation, the receivers are assumed to have equal gain, which may not be the case in practice.

The noise in each frequency channel of an FFT is then Rayleigh-distributed in amplitude, with scale parameter $\sigma / \sqrt{2}$, and uniformly distributed in phase \citep{Papoulis:2002}.
Therefore, denoting the random variable for the noise amplitude as $b$, the complex amplitude measured at two antennas can be written as
\begin{eqnarray}
z_1 & = & a + b\; \e^{i\, \phi_1} \\
z_2 & = & a\; \e^{i\, \theta} + c\; \e^{i\, \phi_2}.
\end{eqnarray}
where $S^2 = a^2 / \E(b^2)$. Here, $\E(\cdot)$ denotes expectation value, and $\theta$ is the phase difference of the transmitter signal across the two antennas.
As the noise phases are uniform-random, and the following analysis is circular-symmetric, the transmitter signal phase difference $\theta$ can be omitted. For this analysis, the preferential direction of the random walk is then along the real axis.

Accumulating the phase variance $s_{1,2}$ for antenna indices $1$ and $2$ as in Eq.~\ref{eq:phasevariance}, corresponds to taking an average of the signal over all data blocks as follows:
\begin{equation}\label{eq:phasevariance_complex}
s_{1,2} = 1 - \frac{1}{N_{\rm blk}} \left| \sum_{N_{\rm blk}} \frac{z_1\, z_2^*}{|z_1| |z_2|} \right|.
\end{equation}

As a first step, we calculate the expected value of the bias in the random walk.  This follows from the expectation value of the real part of the fraction in Eq.~\ref{eq:phasevariance_complex}.
As $b$ and $c$ are independent and identically Rayleigh-distributed, this expectation value is given by
\begin{multline}
\frac{1}{(2 \pi)^2} \int_{-\pi}^{\pi} d\phi_1 d\phi_2 \int_0^\infty db\,dc \frac{2\,b}{\sigma^2}\e^{-b^2/\sigma^2}\, \frac{2\,c}{\sigma^2} \e^{-c^2 / \sigma^2} \\
\frac{{\mathrm{Re}} \left(a^2 + a\,c\,\e^{-i\,\phi_2} + a\,b\,\e^{i \phi_1} + b\,c\,\e^{i(\phi_1 - \phi_2)}\right)}{\sqrt{a^2 + b^2 + 2\,a\,b \cos(\phi_1)}\; \sqrt{a^2 + c^2 + 2\,a\,c \cos(\phi_2)}}.
\end{multline}
As we are dealing with low-amplitude thresholds well below the noise level (i.e.~$S \ll 1$), an asymptotic lowest-order expansion in $a/b$ is used in order to make the integral more tractable.

After collecting the lowest-order terms, the integral evaluates to
\begin{equation}
\E \left({\frac{{\rm Re}\,(z_1 z_2^*)}{|z_1||z_2|}}\right) = \frac{\pi}{4}\, S^2 + \mathcal{O}(S^4).
\end{equation}

The bias $B$ in a random walk of $N_{\rm blk}$ steps is therefore expected to be
\begin{equation}\label{eq:bias}
B = \frac{\pi}{4} S^2 N_{\rm blk}, 
\end{equation}
and the random walk effectively reduces, again to lowest order in $S$, to an unbiased random walk with respect to a point at distance $B$ from the origin.
Using the Rayleigh distribution for the unbiased random-walk distance to the origin, and displacing it by the bias, we obtain for the expected distance:

\begin{multline}
\E(d) = \frac{1}{2\pi}\int_{-\pi}^{\pi}d\phi \int_0^{\infty}dR \;\frac{R}{\tau^2} \;\e^{-R^2/(2\tau^2)} \\ 
\sqrt{R^2 + B^2 + 2\,R\,B\, \cos(\phi)},
\end{multline}
with scale parameter $\tau=\sqrt{N_{\rm blk}/2}$.
To lowest order in $B$ this yields
\begin{equation}\label{eq:excessdistance}
\E(d) \sim \E(d)_{\rm unbiased} + \frac{1}{4} \,B^2 \,\sqrt{\frac{\pi}{N_{\rm blk}}}.
\end{equation}
The excess distance needs to be above a chosen factor $k$ times the standard error of the unbiased random walk distance (see discussion of Eq.~\ref{eq:phase_noiselevel}), using the ensemble having one random walk for each of the $N_{\rm ant}(N_{\rm ant} - 1) / 2$ antenna pairs.
Hence we have a condition
\begin{equation}
\frac{1}{4} \,B^2 \,\sqrt{\frac{\pi}{N_{\rm blk}}} > k\,\beta\, \sqrt{2}\;N_{\rm blk}^{1/2}\; N_{\rm ant}^{-1},
\end{equation}
where the right-hand side is $k$ times the standard error for large $N_{\rm ant}$, approximating the number of antenna pairs by $N_{\rm ant}^2 / 2$.

Comparing these using Eq.~\ref{eq:bias} gives as a threshold, for large $N_{\rm ant}$:
\begin{equation}
S^2 > \frac{8}{\pi} \,\left(\frac{2}{\pi} - \frac{1}{2}\right)^{1/4}\,\sqrt{k}\;N_{\rm blk}^{-1/2}\;N_{\rm ant}^{-1/2},
\end{equation}
reducing to 
\begin{equation}
S^2 > 3.8\;\sqrt{k/6}\;N_{\rm blk}^{-1/2}\;N_{\rm ant}^{-1/2},
\end{equation}
aimed at a 6-sigma detection threshold ($k=6$).

\end{appendix}

\bibliographystyle{aa}
\bibliography{time_calibration}

\end{document}